\documentclass{article}
\usepackage{amsmath}
\usepackage{amssymb}
\usepackage{amsfonts}
\usepackage{a4wide}
\usepackage{amsthm}
\usepackage{tikz}
\usepackage{float}

\newcounter{NN}
\newtheorem{example}[NN]{Example}

\title{\bfseries%
  Three classes of quadratic vector fields for which the Kahan discretization is the root of a generalised Manin transformation
}
\author{%
  Peter H. van der Kamp$^1$, Elena Celledoni$^2$, Robert I. McLachlan $^3$, David I. McLaren$^1$, \\
  Brynjulf Owren$^2$, G.R.W. Quispel$^1$.\\[12pt]
\normalsize
  1~La Trobe University, Victoria 3086, Australia\\
  \normalsize
  2~Norwegian University of Science and Technology, 7491, Trondheim, Norway\\
  \normalsize
  \normalsize
  3~Massey University, Palmerston North, New Zealand\\
\normalsize
}
\date{}

\def\X{{\mathbf x}}
\def\p{{\mathbf p}}

\def\b{{\mathbf b}}
\def\r{{\mathbf r}}
\def\s{{\mathbf s}}
\def\v{{\mathbf v}}
\def\w{{\mathbf w}}
\def\D{{N}}

\newcommand{\cJ}{\mathcal{J}}

\begin{document}

\maketitle
\pagestyle{empty}
\thispagestyle{empty}

\begin{abstract}
We apply Kahan’s discretisation method to three classes of 2-dimensional quadratic vector fields with quadratic, resp cubic, resp quartic Hamiltonians.
We show that the maps obtained in this way can be geometrically understood as the composition of two involutions, one of which is a (linear) symmetry switch, and the other is a generalised Manin involution.
\end{abstract}
\section{Introduction}
\label{sec:INTRODUCTION}

Kahan's method for discretizing quadratic differential equations was introduced in \cite{Kahan}. It was rediscovered in the context of integrable systems by Hirota and Kimura \cite{HirKim}. Suris and collaborators extended the applications to integrable systems significantly in a series of papers \cite{PetSur1}, \cite{PetSur2}, \cite{PetSur3}, \cite{PetPfaSur}, \cite{HonePet}. Applications to non-integrable Hamiltonian systems and the use of polarisation to discretise arbitrary degree Hamiltonian systems were studied in \cite{celledoni1}, \cite{celledoni2} and \cite{celledoni3}. Two classes of $2$-dimensional ODE systems of quadratic vector fields where the Kahan discretization is integrable were presented in \cite{CMMOQ}. The latter systems are of the form
 \begin{equation}
 \label{eq:1}
 \frac{d\X}{dt} = \varphi(\X) \cJ \nabla H(\X),
 \end{equation}
 where
 $$\X:=(x,y),\qquad \cJ:= \left( \begin{array}{cc}
0 & 1  \\
-1 & 0  \end{array} \right),$$
and $\varphi(\X)$ and $H(\X)$ are scalar functions of the components of $\X$. In the present paper we show that for one of these classes, and for two other classes, the Kahan map, which for homogeneous quadratic vectorfields $\frac{dx_i}{dt}=\sum_{j,k}a_{ijk}x_jx_k$ is defined by
$\frac{x_i^\prime-x_i}{h}=\sum_{j,k}a_{ijk}(x^\prime_jx_k+x_jx^\prime_k)/2$,
can be geometrically understood as the composition of two involutions, one of which is a symmetry switch and the other is a generalised Manin involution, both introduced in \cite{KMQ}. This implies that in each case the Kahan map is the root of a generalised Manin transformation, and hence that there is a (fractional affine) transformation which brings the map into symmetric QRT-form \cite{QRT1,QRT2}.

\section{Generalised Manin involutions and symmetry switches}
Let $\p$ be a base point of a pencil of cubic curves $\alpha F_a(\X) + \beta F_b(\X) =0$, i.e. we have $F_a(\p)=F_b(\p)=0$. A Manin involution, $\iota_\p$, maps a point $\r$ to the point $\s=\iota_\p(\r)$ uniquely given by the third intersection of the line $\p\r$ and the curve of the pencil that contains $\r$ \cite{Dui,Man}. We call $\iota_\p$ a $\p$-switch, and the point $\p$ its involution point. A Manin transformation is the composition of two Manin involutions. A generalised Manin involution \cite{KMQ} preserves a pencil of degree $\D$, where $\D$ is not necessarily 3. When $\D=2$ the involution point $p$ can be chosen arbitrarily, for $\D>3$ the degree $\D$ pencil should have a base point which is a singular point of multiplicity $\D-2$. It was shown in \cite{KMQ} that it suffices to consider pencils of degree $\D<5$ and that a generalised Manin transformation can be written in QRT-form by a projective collineation. A transformation $\sigma$ is called a symmetry switch of the pencil $P=0$ if $\sigma$ is a symmetry of $P$ and it is an involution. The following result was proven in \cite{KMQ}. Let $\sigma$ be a symmetry switch of a pencil $P=0$ which maps lines to lines (so it is a projective collineation). Then
\[
\tau_\p = \iota_{\sigma(\p)}\circ \iota_\p = \rho_\p^2,\quad \text{ with } \rho_\p=\sigma \circ \iota_\p=\iota_{\sigma(\p)} \circ \sigma.
\]
The map $\rho_\p$ is called the root of $\tau_\p$.

\section{Concomitants of linear and quadratic forms}
We define linear and quadratic forms
\[
L=L(\X):=ax+by,\qquad Q=Q(\X):=cx^2+2dxy+ey^2.
\]
The three classes of quadratic vectorfields we consider are of the form \eqref{eq:1}
with $\varphi(\X)=L^{2-i}$ and $H(\X)=L^{i-1}Q$ and $i=1,2,3$.
All relevant quantities, e.g. modified Hamiltonian for the Kahan map
and involution point for the Manin involutions will be given in terms of the concomitants (i.e. invariants, covariants, symmetry) defined here, cf. \cite[Page 252]{Ell}.

Let $\eta$ be an element of $SL(2)$ acting on $\X$. This induces an action of $SL(2)$ on the coefficients $a,b,c,d,e$ which we denote by $\eta^\prime$. The discriminant of $Q$,
\[
D:=ce-d^2
\]
and the eliminant (resultant of $L$ and $Q$),
\[
E:=2abd-a^2e-b^2c,
\]
are invariants and the Jacobian determinant $\partial(L,Q)/\partial(x,y)$,
\[
G=G(\X):=(ad-bc)x+(ae-bd)y,
\]
(which is the harmonic conjugate of $L$ with respect to $Q$) is covariant, i.e.
\[
\eta^\prime(D)=D,\quad \eta^\prime(E)=E,\quad \eta^\prime(G)=\eta(G).
\]
In terms of
\[
\v:=(b,-a),\quad \w:=(ad-bc,ae-bd)
\]
we have $G = \X \cdot \w$ and $E=G(\v)$.

A particular linear symmetry switch, introduced in \cite{KMQ}, is relevant here. We define
\begin{equation} \label{lsw}
\sigma_{a,b,c,d,e}:\X\rightarrow \X-\frac{2G(\X)}{E}\v.
\end{equation}
A special case of $\sigma$ is $\sigma_{a,a,c,d,c}(\X)=(y,x)$ and the matrices of $\sigma_{a,a,c,d,c}$ and $\sigma_{a,b,c,d,e}$ are conjugate. In the sequel we will omit the index $_{a,b,c,d,e}$. Geometrically, the linear transformation $\sigma$ given by (\ref{lsw}) is a reflection in the line through $(0,0)$ perpendicular to $\w$ along a line with direction $\v$, i.e. we have
\[
\sigma(\v)=-\v,\qquad \sigma(\cJ \w)=\cJ \w.
\]
Importantly, $\sigma$ (\ref{lsw}) leaves the forms $L$ and $Q$ invariant (and it also negates the linear form $G$), that is
\[
L(\sigma(\X))=L(\X),\qquad Q(\sigma(\X))=Q(\X), \qquad G(\sigma(\X))=-G(\X).
\]

\section{A quadratic Hamiltonian in $2$D} \label{sqh}
Consider the $2$-dimensional Suslov system \eqref{eq:1} where $\varphi(\X)=L$, and the homogeneous Hamiltonian has the form  $H=H(\X)=Q$. The Kahan map for this system is explicitly given by
\begin{equation} \label{quadramap}
\kappa_1(\X) = \frac{\X-h(G\X-L\cJ \nabla H)}{1-hG+2h^2DL^2}.
\end{equation}
It preserves the modified Hamiltonian $\widetilde{H}(\X)= Q/T$ with $T=T(\X)=1+h^2DL^2$, cf. \cite[Eq. (18)]{celledoni2}, and it is measure preserving with density
\begin{equation}
\label{eq:density}
\frac{1}{LQ},
\end{equation}
cf. \cite[Eq. (17)]{celledoni2}.
The map (\ref{quadramap}) can be written as a composition $\kappa_1=\sigma \circ \iota_\b$, where $\sigma$ is given by (\ref{lsw}) and
\[
\iota_\b(\X)=\X+(1-\frac{1+hG}{1-hG+2h^2DL^2})(\b-\X)
\]
where
\begin{equation} \label{B}
\b=\frac{\v}{hE}.
\end{equation}
The map $\iota_\b$ is the generalised Manin involution with involution point $\b$, cf. \cite[equation (2)]{KMQ} with $\D=2$ and $F_a=Q$ and $F_b=T$. The point $\b$ is the intersection point of the lines $L=0$ and $1-hG=0$,
and we have $\sigma(\b)=-\b$.

\begin{example}
We take $a=e=-c=1$, $b=d=0$. Then $L=x$ and $Q=y^2-x^2$,
$\widetilde{H}=Q/T$ with $T=1-h^2x^2$,
and
\[
\kappa_1(\X)=\left(\frac{x(1+hy)}{1-hy-2h^2x^2},
\frac{2hx^2-hy^2+y}{1-hy-2h^2x^2}\right).
\]
The linear involution is $\sigma(\X)=(x,-y)$.
The curves $Q=0$ and $T=0$ intersect in four points, namely
$(\epsilon,\delta)/h$ with $\epsilon,\delta\in\{\pm1\}$.
The involution point is $\b=(0,1/h)$, which is not one of the base points. We choose $h=\frac12$ and have drawn three level sets of the modified Hamiltonian in Figure \ref{F3}. We have also plotted the images of $(1/3,1)$, $(1,2)$ and $(3,0)$ under the Manin involution
\[
\iota_\b(\X)=\sigma \circ \kappa_1(\X)=\left(\frac{x(1+hy)}{1-hy-2h^2x^2},
-\frac{2hx^2-hy^2+y}{1-hy-2h^2x^2}\right).
\]
\begin{figure}[h]
\begin{center}
\includegraphics[width=10cm,height=6cm]{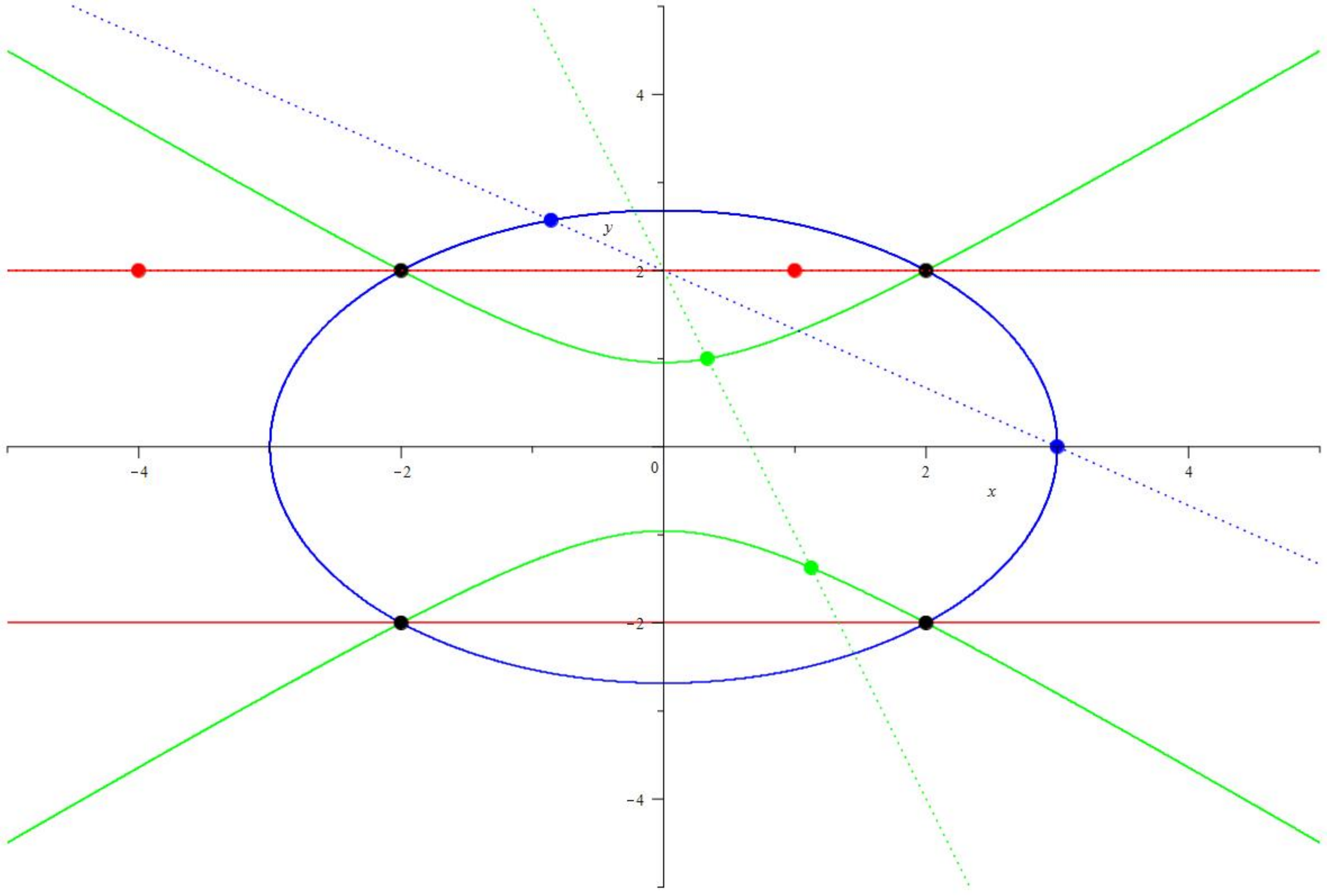}
\caption{\label{F2} The curves $\widetilde{H}=\frac{32}{35}$, $\widetilde{H}=4$,
$\widetilde{H}=\frac{36}{5}$, in resp. green, red and blue.}
\end{center}
\end{figure}
\end{example}

The projective collineation $\pi:(x,y)\rightarrow (u,v)=(1+hG,1-hG)/L$ brings the point $\b$ to the point at infinity $(\infty,0)$. Hence the $\b$-switch is  transformed into the horizontal switch $\overline{\iota_\b}=\pi\circ\iota_\b\circ\pi^{-1}=\iota_1$,
where, in terms of $k=4h^2D$,
\[
\iota_1:(u,v)\rightarrow \left(\frac{v^2+k}{u},v\right).
\]
The symmetry switch $\sigma$ is transformed into the standard symmetry switch $\overline{\sigma}:(u,v)\rightarrow (v,u)$, and thus the map $\kappa_1$ is brought into QRT form
\begin{equation} \label{qrt1}
\overline{\kappa_1} = \overline{\sigma} \circ \iota_1: (u,v)\rightarrow \left(v,\frac{v^2+k}{u}\right).
\end{equation}
The modified Hamiltonian transforms into
\[
\overline{Q/T}=\frac{(u-v)^2+k}{(u+v)^2+k},
\]
which is invariant under (\ref{qrt1}).

\section{A cubic Hamiltonian in $2$D}
Next we consider the $2$-dimensional ODE system \eqref{eq:1} where $\varphi(\X)=1$, and the homogeneous Hamiltonian has the form  $H=
H(\X)=LQ$. The Kahan map for this system,
\begin{equation} \label{ceq}
\kappa_2(\X) =\frac{\X + h \cJ \nabla H}{R}, \qquad R=R(\X)=1+h^2(3DL^2-G^2),
\end{equation}
preserves the modified Hamiltonian $\widetilde{H}=H/R$, cf. \cite[Eq. (4)]{celledoni2}.
The map is measure preserving with density (\ref{eq:density}) and it is the composition of $\sigma$ (\ref{lsw}) and the Manin map
\[
\iota_\b=\X+Z.(\b-\X)
\]
where $\b$ is given by (\ref{B}) and
\[
Z=1-\frac{1+2hG}{R}
\]
can be obtained from \cite[equation (2)]{KMQ} with $\D=3$ and $F_a=LQ$ and $F_b=R$. Note that here the involution point $\b$ is a base point, of the pencil $\alpha H+\beta R=0$, as it is the intersection of the lines $L=0$ and $hG=1$.

\begin{example}
We take $a=e=-c=1$, $b=d=0$. Then $H=x(y^2-x^2)$,
\[
\widetilde{H}=\frac{x(y^2-x^2)}{1-(3x^2+y^2)h^2},
\]
and
\[
\kappa_2(\X)=\left(\frac{x(1+2hy)}{1-h^2(3x^2+y^2)},
\frac{(3x^2-y^2)h+y}{1-h^2(3x^2+y^2)}\right).
\]
The linear involution is $\sigma(\X)=(x,-y)$.
The curves $H=0$ and $R=0$ intersect in six points, namely
\[
\left(0, \frac{\epsilon}{h}\right), \qquad \left(\frac{\epsilon}{2h},\frac{\delta}{2h}\right), \qquad \epsilon,\delta\in\{\pm1\}.
\]
The involution point is $\b=(0,1/h)$. Choosing $h=1/2$ we have drawn three level sets of the modified Hamiltonian in Figure \ref{F3}, where we have also indicated the images of $(2,3)$, $(1,2)$ and $(3,0)$ under the Manin involution $\iota_\b=\sigma\circ\kappa_2$.
\begin{figure}[h]
\begin{center}
\includegraphics[width=10cm,height=6cm]{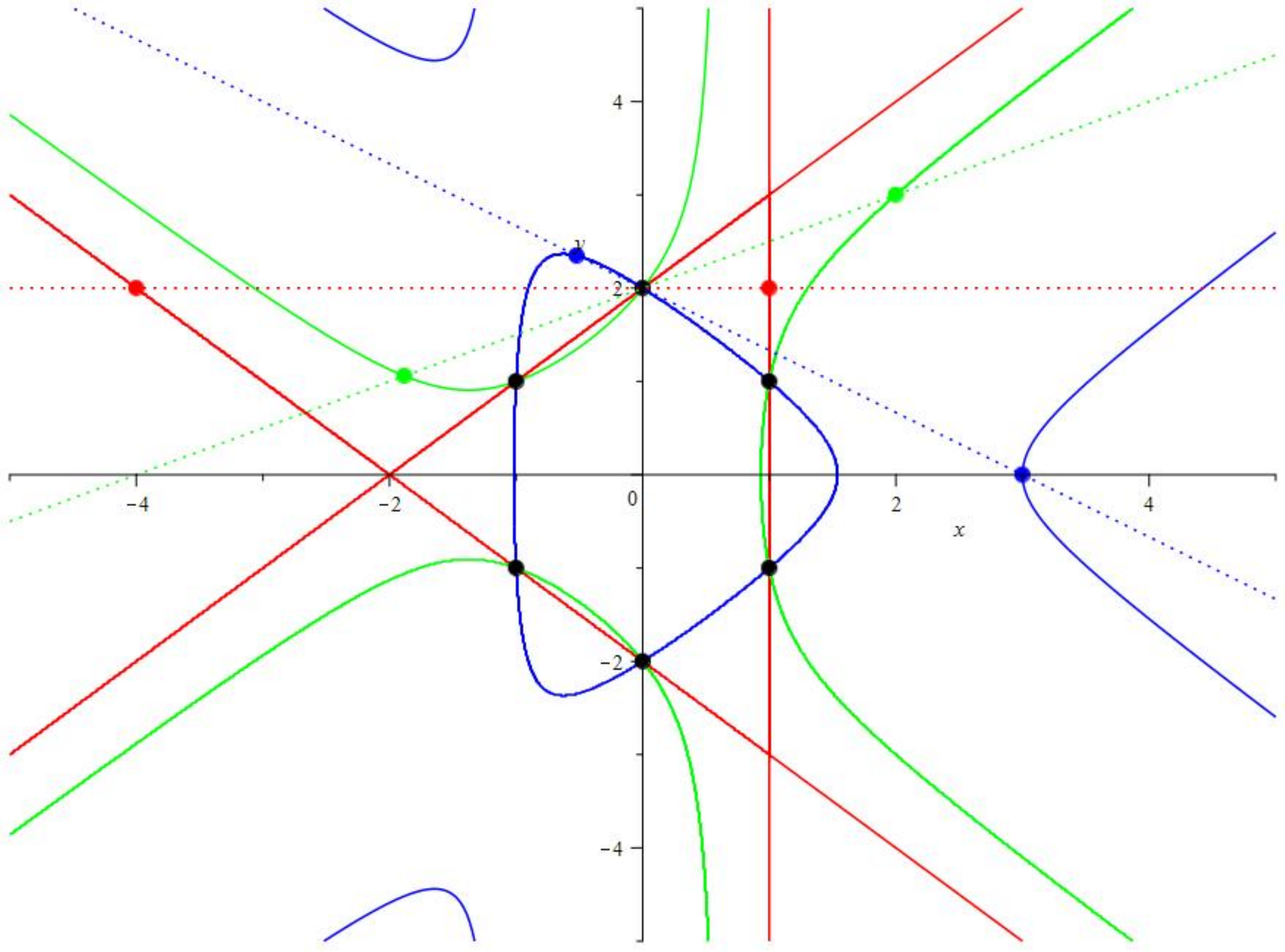}
\caption{\label{F3} The curves $\widetilde{H}=-\frac{40}{17}$, $\widetilde{H}=-4$,
$\widetilde{H}=\frac{108}{23}$, in resp. green, red and blue.}
\end{center}
\end{figure}
\end{example}

The projective collineation $\pi$ from section \ref{sqh} brings the map (\ref{ceq}) in QRT form,
\[
\overline{\kappa_2} : (u,v)\rightarrow \left(v,\frac{(u+v)v+3k}{3u-v} \right),
\]
which leaves
\[
\overline{H/R}=\frac{(u-v)^2+k}{(u+v)(2uv+\frac 32k)}
\]
invariant.
\section{A quartic Hamiltonian in $2$D}
Consider the $2$-dimensional ODE system \eqref{eq:1} where $\varphi(\X)=\frac{1}{L}$,
and the homogeneous Hamiltonian has the form $H(\X)=L^2Q$. Then the Kahan map for this system,
\begin{equation} \label{qeq}
\kappa_3(\X)=\frac{\X+h(G\X+L^{-1}\cJ\nabla H)}{(1-hG)(1+2hG)+4h^2DL^2}
\end{equation}
preserves the modified Hamiltonian $
\widetilde{H}(\X)=\frac{H}{S}$ with $S=S(\X)=(1-h^2G^2)(1+h^2(8DL^2-G^2))$
and it is measure preserving with density (\ref{eq:density}),
cf. \cite[Section 2]{CMMOQ}. It is the composition of $\sigma$, as defined by (\ref{lsw}), and the Manin map
\[
\iota_\b=\X+(1-\frac{1+3hG}{(1-hG)(1+2hG)+4h^2DL^2})(\b-\X)
\]
where $\b$ is again given by (\ref{B}).
This formula agrees with \cite[equation (2)]{KMQ} taking $\D=4$ and $F_a=L^2Q$ and $F_b=S$. The involution point $\b$ is a double base point, as $\b$ is also on the curve $h^2EQ=1$.

\begin{example}
Taking $a=e=-c=1$, $b=d=0$ yields $H=x^2(y^2-x^2)$,
\[
\widetilde{H}=\frac{H}{S}=\frac{x(y^2-x^2)}{(1-h^2y^2)(1-h^2(8x^2+y^2))},
\]
and
\[
\kappa_3(\X)=\left(\frac{x(1+3hy)}{1+hy-2h^2(y^2+2x^2)},
\frac{y+h(4x^2-y^2)}{1+hy-2h^2(y^2+2x^2)}\right).
\]
The linear involution is $\sigma(\X)=(x,-y)$.
The curves $H=0$ and $S=0$ intersect in 10 points,
\[
\left(0, \frac{\epsilon}{h}\right), \qquad
\left(\frac{\epsilon}{h},\frac{\delta}{h}\right), \qquad
\left(\frac{\epsilon}{3h},\frac{\delta}{3h}\right), \qquad \epsilon,\delta\in\{\pm1\}.
\]
The involution point is $\b=(0,1/h)$. Taking $h=1/3$ we have drawn three level sets of the modified Hamiltonian in Figure \ref{F4}, as well as the points $(-2,2)$, $(\frac12,2)$, $(3,2)$ and their images under the Manin involution $\iota_\b=\sigma\circ\kappa_3$.
\begin{figure}[h]
\begin{center}
\includegraphics[width=10cm,height=6cm]{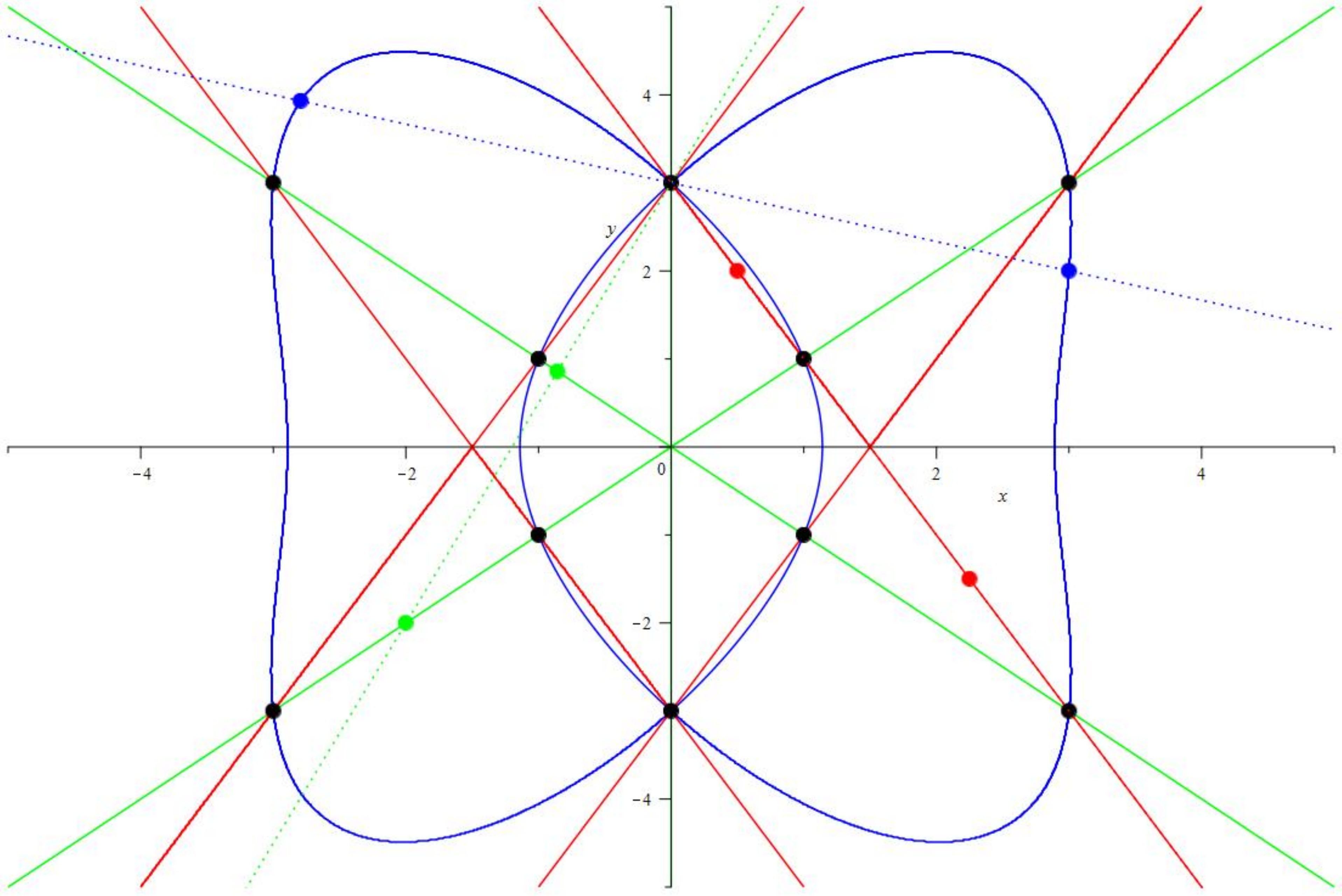}
\caption{\label{F4} The curves $\widetilde{H}=0$, $\widetilde{H}=\frac{81}{16}$,
$\widetilde{H}=\frac{729}{67}$, in resp. green, red and blue.}
\end{center}
\end{figure}
\end{example}

The projective collineation $\pi$ from section \ref{sqh} brings the map (\ref{qeq}) in QRT form,
\[
\overline{\kappa_3} : (u,v)\rightarrow \left(v,\frac{uv+2k}{2u-v} \right),
\]
which leaves
\[
\overline{H/S}=\frac{(u-v)^2+k}{4uv(uv+2k)}
\]
invariant.

\section{Summary}
We have shown that the Kahan discretization of the ODE (\ref{eq:1}) with $\varphi(\X)=L^{2-i}$ and $H(\X)=L^{i-1}Q$ for each $i=1,2,3$ takes the form $\kappa=\sigma \circ \iota_\b$ where $\iota_\b$ is the $\b$-switch with involution point $\b=\v/(hE)$, and $\sigma$ is both a linear map and a symmetry of the preserved pencil which has degree $i+1$. Therefore, in each case the Kahan map is the root of the generalised Manin transformation $\tau_\b = \iota_{\sigma(\b)} \circ \iota_\b$. According to \cite{KMQ} a generalised Manin involution $\iota_\p$ which preserves a pencil $\alpha F_a(\X) + \beta F_b(\X) =0$ of degree $2\leq \D\leq 4$ is measure preserving with density $L^{\D-3}/F_a$, where $L$ is any line through $\p$. This implies, as we have $F_a=H$, that the density of the measure preserved by the Kahan map is the same for each $i$, namely $1/(LQ)$. For each Kahan map we have provided its symmetric QRT form.

\section*{Acknowledgment}

This work was supported by the Australian Research Council, by the Research Council of Norway, by the Marsden Fund of the Royal Society of New Zealand,
and by the European Union's Horizon 2020 research and innovation programme under the Marie Sk\l{}odowska-Curie grant agreement No. 691070.


\end{document}